\documentclass[aps,prd,amsmath,amssymb]{revtex4}

\newcommand{\beq}{\begin{equation} }
\newcommand{\eeq} {\end{equation} }
\newcommand{\bed}{\begin{displaymath} }
\newcommand{\eed} {\end{displaymath} }
\begin{document}

\title{Stabilizing the axion and a natural solution to the $\mu$ problem of supersymmetry\footnote{\uppercase{T}alk presented at {\it \uppercase{SUSY} 2003: \uppercase{S}upersymmetry in
the \uppercase{D}esert}\/, held at the \uppercase{U}niversity of
\uppercase{A}rizona, \uppercase{T}ucson, \uppercase{AZ},
\uppercase{J}une 5-10, 2003. \uppercase{T}o appear in the
\uppercase{P}roceedings.}}

\author{Kai Wang}
\email{wangkai@hep.phy.okstate.edu}
\affiliation{Department of Physics, Oklahoma State University\\
145 Physical Sciences II\\ Stillwater, OK 74078-3072}


\begin{abstract}The axion solution to the strong CP problem makes use
of a global Peccei-Quinn (PQ) $U(1)$ symmetry which is susceptible
to violations from quantum gravitational effects. We show
explicitly how discrete gauge symmetries can protect the axion
from such violations. PQ symmetry emerges as an accidental global
symmetry from discrete gauge symmetries which are subgroups of the
anomalous $U(1)$ of string origin. We also show how the
Dine-Fischler-Srednicki-Zhitnitsky (DFSZ) axion model provides a
natural solution to $\mu$ problem of supersymmetry as $\mu\sim
M_{\rm SUSY}\sim M^2_{\rm PQ}/M_{\rm Pl}$.
\end{abstract}
\maketitle



\section{The Strong CP Problem and Stabilizing the Axion}

CP violation (CPV) can exist in the QCD Lagrangian arising from
the instanton induced Chern-Simons type gluon-gluon coupling \bed
{\mathcal L}\supset\theta g^2_s
\epsilon^{\mu\nu\sigma\rho}G^{\alpha}_{\mu \nu}
G^{\alpha}_{\sigma\rho}/{64\pi^2}=\theta g^2_s G^{\alpha}_{\mu
\nu} \tilde{G}^{\alpha\mu\nu}/{32\pi^2}.\eed In addition, there is
another CPV source from the quark mass matrices. This results in
an observable parameter $\bar{\theta}$ defined as \bed
\bar{\theta}=\theta+\mathrm{arg}(\mathrm{det}M_U~
\mathrm{det}M_D).\eed Such a $\bar{\theta}$ would lead to a
neutron electric dipole moment (EDM) of order $d_n\simeq
5\times10^{-16}~\bar{\theta}~\rm{e cm}$, while the current
experiment limit is $d_n<10^{-25}~\rm{e cm}$. This puts a strong
constraint, $\bar{\theta}< 10^{-10}$. Peccei-Quinn (PQ)
symmetry\cite{PQ} is an elegant solution to this so-called strong
CP problem. It introduce a global $U(1)$ symmetry, broken by the
QCD anomaly, which generates a pseudo-Goldstone particle $a$, the
axion. Non-perturbative effect then induces a term in the
Lagrangian \bed {\mathcal L}\supset (a/f_a) g^2_s G^{\alpha}_{\mu
\nu} \tilde{G}^{\alpha\mu\nu}/{32\pi^2}.\eed $\overline{\theta}$
is then promoted to this dynamical field axion as $a(x)/f_a$.
Minimizing the axion potential \bed  V(a) \propto \Lambda^4_{\rm
QCD}(1-\cos(a(x)/f_a)),\eed consequently $\bar{\theta}=\langle
a\rangle/f_a=0$.\footnote{Due to the periodicity of the potential,
$\langle a\rangle=2n\pi f_a$. Some detailed discussion can be
found in various review papers listed as references in \cite{wk}.
} The strong CP problem is then solved.
 $f_a$
is the (model dependent) axion decay constant\cite{WW} and it is
constrained to be $f_a=(10^{10}-10^{12})~{\rm GeV}$ by the
combined limits from laboratory experiments, astrophysics and
cosmology. Hence, only the ``invisible axion" models, which have
appropriate values of $f_a$, are favored.\cite{dfsz,kim} The
couplings of the axion with the Standard Model (SM) fields are
highly suppressed in these models. Although axion arise as a
pseudo-Goldstone particle when the PQ symmetry is explicitly
broken by its QCD anomaly, the axion can acquire a tiny mass
through higher order non-perturbative effect. The mass of the
axions can be estimated to be \bed m_a\sim \Lambda^2_{\rm
QCD}/f_a\sim 10^{-4}~{\rm eV}.\eed

Quantum gravitational effects can potentially violate the global
PQ symmetry as they can break all global symmetries while
respecting gauge symmetries. In the axion models, a possible
quantum gravity generated non-renormalizable term \bed {\mathcal
L}\supset S^n/M_{\rm Pl}^{n-4}\eed is in principle allowed. This
term would lead to \bed \bar{\theta} \simeq f_a^{n}/({{M_{\rm
Pl}}^{n-4}\Lambda_{QCD}^{4}}).\eed Since both $\bar{\theta}$ and
$f_a$ are highly constrained, $n \geq 10$ is necessary. To avoid
such kind of violations, one solution is to introduce a discrete
gauge symmetry. The PQ symmetry  arises only as an accidental
global symmetry from it. Discrete gauge symmetries\cite{kw} are
remnants of gauge groups after spontaneous symmetry breaking and
they are left intact by quantum gravity. There is no gauge boson
associated with such a symmetry. However, for it to be of gauge
origin, it must be anomaly free.

Conventionally, absence of anomalies complicates the particle
spectrum of axion models. However, the Type I and Type IIB string
theories provide a new candidate that cancels the anomalies
without enlarging the particle content. In the low energy
effective theory of such string theories, there exists one
anomalous $U(1)_A$ symmetry. The Green-Schwarz Mechanism
(GSM)\cite{gs} is effective in cancelling the anomalies. The
anomalous $U(1)_A$ symmetry is broken by a Higgs field
spontaneously near the string scale. If this Higgs has charge $N$
under $U(1)_A$, there will be a remnant $Z_N$ discrete gauge
symmetry. At low energy, one can check if $Z_N$ can be embedded in
to the anomalous $U(1)_A$ by a discrete version of GSM. It
requires \bed (A_{3}+{\rm mod}~N/2)/{k_3}=({A_2}+{\rm
mod}~N/2)/{k_2}=({A_1}+{\rm mod}~N/2)/{k_1},\eed where $A_i$ are
mixed anomaly coefficients with the SM gauge group $G_i$ as
$A_{[G_i]^2\times Z_N}$ and $k_i$ are the corresponding Kac-Moody
levels. Notice under a $Z_N$ symmetry, the anomaly coefficient can
differ by ${\rm mod}~N/2$ since one can always have vectorial
particles in the theory which do not contribute to the anomalies
at the $U(1)$ level.

Here we take the supersymmetric (SUSY) DFSZ axion model as an
explicit example. The superpotential of the DFSZ axion model
contains a term $\lambda {H_u}H_{d}S^{2}/M_{\rm Pl}$. After $H_u$,
$H_d$ and $S$ develop VEVs, the global PQ symmetry is broken and
the axion arises as a pseudo-Goldstone particle. Since the
superpotential is holomorphic, one cannot write ${S^\dagger}^2
S^2$ type term. In addition to the $S$ field, another singlet
$\tilde{S}$ is needed so that the axion is invisible and at the
same time, PQ can be broken. The superpotential of the model now
is \bed W\supset \lambda_1 H_u H_d S^2/M_{\rm Pl}+ \lambda_2 S^2
\tilde{S}^2/M_{\rm Pl}.\eed One explicit example of $Z_{22}$
discrete gauge symmetry is given. The charge assignment under
$Z_{22}$ is listed as \bed \{Q=3, ~u^{c}=19, ~d^{c}=1, ~L=11,
~e^{c}=15, ~\nu^{c}=11, ~H_u=22, ~H_d=18, ~S=13, ~\tilde{S}=20
\}.\eed The mixed anomalies are  $\{A_2=6,
   A_3=17\}$. It apparently satisfies the GSM condition. $S^{22}/M^{19}_{\rm Pl}$ is the leading allowed term in the
superpotential due to potential quantum gravity correction, which
only induces $\bar{\theta}\lesssim 10^{-130}$.

In this model, the $R$-parity is not automatic, for instance, $L
H_u S \tilde{S}$ is allowed. To get an exact $R$-parity, one can
introduce an additional $Z_2$ where all the SM matter fields are
odd but $H_u$, $H_d$, $S$ and $\tilde{S}$ are even. This is the
unbroken subgroup of the gauge symmetry $U(1)_{\rm B-L}$ even with
the presence of Majarona neutrino mass term.

The KSVZ Axion model\cite{kim}, can also be stabilized by discrete
gauge symmetries.\cite{wk} It requires extra matter fields which
are vectorial under QCD. An interesting case occurs when the
matter fields include two fields and together form a fundamental
representation of the $SU(5)$ GUT group. The theory is then
automatically consistent with GSM, while providing nonzero QCD
anomaly.

\section{DFSZ Axion: A Natural Solution to the $\mu$ Problem}

The $\mu$ problem is an intriguing puzzle of SUSY extension of the
SM where $\mu$ is the Higgs mass parameter in \bed W\supset \mu
H_u H_d \eed. Why should $\mu$ be of order the soft SUSY breaking
mass scale $M_{\rm SUSY}$, rather than the Planck scale $M_{\rm
Pl}$. There are two well-known attempts to solve this problem. One
is through a non-renormalziable K\"{a}hler potential, \bed
{\mathcal L}\supset \int d^4 \theta H_u H_d Z^*/M_{\rm Pl},\eed
namely the Giudice-Masiero mechanism with an additional
R-symmetry. Another one is the NMSSM models with a global $Z_3$
symmetry at low energy or a $U(1)'$ gauge symmetry broken near
$M_{\rm SUSY}$.

Imposing a new physics scale $M_{\rm PQ}~(
f_a=(10^{10}-10^{12})~{\rm GeV})$, the axion models provide
another approach to the $\mu$ problem\cite{wk,mu} as \bed \mu\sim
M^2_{\rm PQ}/M_{\rm Pl}.\eed In the case of DFSZ axion model, a
$\mu$ term automatically arises after PQ symmetry breaks.

The question now is how to naturally understand the origin of
$M_{\rm PQ}$ from a higher energy theory. It is interesting that
in the SUGRA mediated SUSY breaking models, one also has to impose
a new physics scale of order ${\cal O}(10^{11}~{\rm GeV})$. In
these models, this intermediate scale can be generated
dynamically. Practically, this intermediate scale can then be
identified as $M_{\rm PQ}$. Here we propose a model involving SUSY
breaking.\cite{babu} Having made use of $M_{\rm SUSY}$, this
approach certainly requires that SUSY breaking mediation scale is
greater than $M_{\rm PQ}$. A simple realization of this idea is
the SUGRA model. The superpotential of the model contains \bed
W\supset\lambda_1 H_u H_d S^2/{M_{\rm Pl}} +{\lambda_2
(S\tilde{S})^{2}/{M_{\rm Pl}}}+{S^{22}}/{M^{19}_{\rm Pl}}\eed
which is also consistent with the $Z_{22}$ symmetry in the
previous section. By minimizing the leading-orders potential
including SUSY breaking effects, \bed
V=(\lambda_{2}C{(S\tilde{S})^2}/{M_{\rm Pl}}+h.c)+{m_S}^2|S|^2+{m_
{\tilde{S}}}^2{|\tilde{S}|}^2+4\lambda_2
{|S\tilde{S}|^2}(|S|^2+{|\tilde{S}|}^2)/{M_{\rm Pl}^2},\eed where
$m_S$ and $m_{\tilde{S}}$ are soft breaking masses of order
$M_{\rm SUSY}$, one obtains \bed
f_a^2={C\pm\sqrt{C^2-12{m_{S}}^2}}M_{\rm Pl}/{12\lambda_2}.\eed So
\bed f_a \sim \sqrt{M_{\rm Pl}M_{\rm SUSY}}
 \sim 10^{11}~\mathrm{GeV}.\eed Since
the $F$-component of the field $S$, \bed F_S\sim M_{\rm PQ} M_{\rm
SUSY},\eed the dominant contribution for the $B$ parameter which
appears in the soft bilinear SUSY breaking term \bed {\mathcal
L_{\rm soft}}\supset B\mu H_u H_d\eed arises from the
superpotential $H_u H_d S^2/M_{\rm Pl}$ as \bed B\mu=\langle
S\rangle \langle F_S \rangle /M_{\rm Pl}\sim M^2_{\rm SUSY}.\eed
So it is difficult to distinguish it from the usual MSSM via
electroweak physics. However, as the axion can be a cold dark
matter candidate, one can still distinguish the model in
cosmology. In this model, the two PQ Higgs bosons have masses of
order $M_{\rm SUSY}$ but their mixings with the doublet Higgs are
highly suppressed. The orthogonal combination to the axion
acquires a mass of order $M_{\rm SUSY}$. The axino and saxino
masses are both around $M_{\rm SUSY}$. The axino can mix with the
Higgsino with a tiny mixing angle of order $(M_{\rm SUSY}/M_{\rm
Pl})^{1/2}\sim 10^{-7}$. Therefore, the axino can decay to a
bottom quark and a sbottom squark with a lifetime \bed \tau\sim
10^{-11}~\mathrm{sec}.\eed This is a consistent picture with
big-bang cosmology since the axino decays occur earlier than the
nucleosynthesis era.

\section{Acknowledgement}
The author is very grateful for the other two collaborators of
this project, K.S. Babu and Ilia Gogoladze for useful discussions
and suggestions.

\end{document}